\documentclass[aps,prl,twocolumn]{revtex4}
\usepackage{graphicx}

\begin{document}


\title{Evidence for unconventional superconducting fluctuations in heavy-fermion compound CeNi$_2$Ge$_2$}

\author{S. Kawasaki$^1$}%
\author{T. Sada$^1$}
\author{T. Miyoshi$^2$}
\author{H. Kotegawa$^2$}%
\author{H. Mukuda$^1$}%
\author{Y. Kitaoka$^1$}
\author{T. C. Kobayashi$^2$}
\author{T. Fukuhara$^3$}
\author{K. Maezawa$^3$}
\author{K. M. Itoh$^4$}
\author{E. E. Haller$^5$}

\affiliation{$^1$Department of Materials Engineering Science, Graduate School of Engineering Science, Osaka University, Toyonaka, Osaka 560-8531, Japan\\$^2$Department of Physics, Faculty of Science, Okayama University, Okayama 700-8530, Japan\\$^3$Faculty of Engineering, Toyama Prefectural University, Toyama 939-03, Japan\\$^4$Department of Applied Physics and Physico-Informatics, Keio University, Yokohama 223-8522, Japan\\$^5$Department of Materials Science and Engineering, University of California at Berkeley and Lawrence Berkeley National Laboratory, Berkeley, CA 94720, USA}%


\date{\today}

\begin{abstract}
We present evidence for unconventional superconducting fluctuations in a heavy-fermion compound CeNi$_2$Ge$_2$. The temperature dependence of the $^{73}$Ge nuclear-spin-lattice-relaxation rate $1/T_1$ indicates the development of magnetic correlations and the formation of a Fermi-liquid state at temperatures lower than $T_{\rm FL}=0.4$ K, where $1/T_1T$ is constant. The resistance and $1/T_1T$ measured on an as-grown sample decrease below $T_{\rm c}^{\rm onset} = 0.2$ K and $T_{\rm c}^{\rm NQR} = 0.1$ K, respectively; these are indicative of the onset of superconductivity. However, after annealing the sample to improve its quality, these superconducting signatures disappear. These results are consistent with the emergence of unconventional superconducting fluctuations in close proximity to  a quantum critical point from the superconducting to the normal phase in CeNi$_2$Ge$_2$.
\end{abstract}



\maketitle 

Unconventional superconductivity observed in the vicinity of the antiferromagnetic (AFM) quantum critical point (QCP) has been one of the most important issues in cerium (Ce)-based heavy-fermion (HF) compounds, since it was universally found at the border of antiferromagnetism in CeCu$_2$Si$_2$, CeRh$_2$Si$_2$, CeIn$_3$, CePd$_2$Si$_2$, and CeRhIn$_5$.\cite{Kitaoka} Therefore, it is believed that superconductivity in these compounds is mediated by the magnetic fluctuations induced near the AFM QCP.\cite{Mathur}  Recently, Yuan $et$ $al$. showed that a robust superconducting (SC) phase under pressure in the prototype HF superconductor CeCu$_2$Si$_2$ is divided into two SC domes in CeCu$_2$(Si$_{1-x}$Ge$_x$)$_2$. Since the second SC phase emerges far from the AFM QCP in CeCu$_2$(Si$_{1-x}$Ge$_x$)$_2$, a valence-fluctuation-mediated SC mechanism is proposed for the onset of this SC phase.\cite{Yuan,Miyake} However, thus far, there are few experimental examples for these types of HF superconductivity. Therefore, the mechanism of superconductivity in HF compounds is still under debate from various view points. 

The HF compound CeNi$_2$Ge$_2$ crystallizes in the ThCr$_2$Si$_2$ structure. The measurements of resistivity and specific heat  at low temperatures clearly revealed  non-Fermi-liquid-like behaviors associated with antiferromagnetism; $\Delta\rho\propto T^{n}(n=1.2\sim 1.5)$ and $\Delta C/T\propto\ln{T}$.\cite{Steglich,Grosche,Aoki} 
In fact, a small amount of substitution for the Ni site in Ce(Cu$_{1-x}$Ni$_x$)$_2$Ge$_2$ and Ce(Ni$_{1-x}$Pd$_x$)$_2$Ge$_2$ leads to an AFM order by expanding its lattice volume.\cite{Sparn,Fukuhara,Knebel}  
Thus, CeNi$_2$Ge$_2$ is located near the AFM QCP.  
Remarkably, there exist several reports indicating that resistance becomes zero below $T_{\rm c}\sim 0.2$ K in CeNi$_2$Ge$_2$, suggesting the onset of superconductivity.\cite{Grosche,Gegenwart}  Therefore, it is suggested that magnetic fluctuations with regard to the AFM QCP are responsible for the onset of superconductivity in this compound. However, since superconductivity in this compound strongly depends on the sample preparation method and/or the nominal stoichiometry in the Ni element,\cite{Gegenwart} experiments revealing bulk superconductivity have not been reported to date.  It should be noted that two SC domes have also been reported in CeNi$_2$Ge$_2$ from resistivity measurements under pressure.\cite{Flouquet,Grosche2}
    
In this letter, we report systematic studies of resistivity and $^{73}$Ge nuclear quadrupole resonance (NQR) in an as-grown and an annealed sample of CeNi$_2$Ge$_2$. The NQR spectrum does not reveal any trace of a magnetic order down to 0.03 K. The temperature ($T$) dependence of the nuclear spin-lattice relaxation rate $1/T_1$ indicates the growth of magnetic correlations followed by an emergence of the Fermi-liquid state below $T_{\rm FL}=0.4$ K, where $1/T_1T$ is constant. A decrease in both resistance and $1/T_1T$ are observed below $T_{\rm c}^{\rm onset}=0.2$ K and $T_{\rm c}^{\rm NQR}=0.1$ K in the as-grown sample, respectively. These anomalies are related to the onset of superconductivity because the application of a tiny magnetic field causes $1/T_1T$ to remain constant. On annealing to improve the quality, the anomalies with regard to the onset of superconductivity disappear. Importantly, the signature towards SC transition has been microscopically investigated by the NQR-$T_1$ measurement, which reveals a reduction without a coherence peak in $1/T_1$ just below $T_{\rm c}$.  However, the result that zero resistance is {\it not} observed could be attributed to the SC fluctuations due to the proximity with a SC QCP. 
\begin{figure}[h]
\centering
\includegraphics[width=6cm]{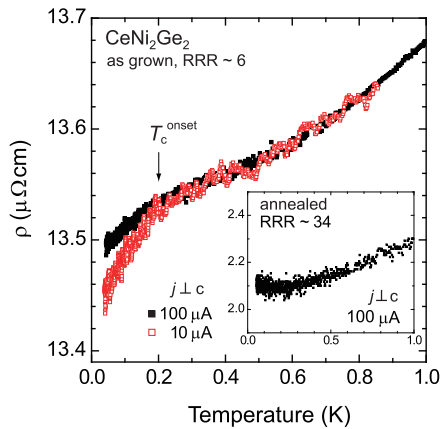}
\caption[]{\footnotesize (Color online) Temperature and current dependences of resistivity for the as-grown sample of CeNi$_2$Ge$_2$. Solid arrow indicates $T_{\rm c}^{\rm onset}$. The inset shows the temperature dependence of resistivity for the annealed sample (see text).}
\end{figure}
 
High-quality single crystals of CeNi$_{2.02}$Ge$_2$ were grown by the Czochralski method and moderately crushed into grains in order to enable easy penetration of the rf pulses into the samples. However, to avoid crystal distortions, the size of the grains is kept larger than 100 $\mu$m. Note that the onset of superconductivity in this compound is extremely sensitive to the sample preparation method and/or the nominal stoichiometry in the Ni element. The sharpest SC transition was reported for the Ni-rich sample CeNi$_{2.02}$Ge$_2$.\cite{Gegenwart}  The measurements of resistance and $^{73}$Ge NQR with nuclear spin $I$ = 9/2 were performed using an almost 100\% $^{73}$Ge enriched sample. A single crystal is used for the resistivity measurement. The $T$ dependence of resistivity was measured in a zero field ($H$ = 0) by the conventional four-probe method. The current flows perpendicular to the $c$ axis. The $T$ dependences of $^{73}$Ge-NQR spectra and $1/T_1$ were measured at $H = 0$ down to $T=0.03$ K using a $^3$He/$^4$He dilution refrigerator. In order to detect the possible onset of some magnetic ordering at low temperatures, the NQR spectrum for the 1$\nu_{\rm Q}$ $(\pm 1/2\leftrightarrow \pm 3/2)$ transition was precisely measured by the Fourier transform method for spin-echo signals.   
 
Figure 1 and the inset show the $T$ dependences of resistance for the as-grown and annealed single crystals, respectively. Note that the resistance at the $ab$-plane decreases for the as-grown sample below $T_{\rm c}^{\rm onset} = 0.2$ K,  but it does not drop to zero; it is consistent with the previous results.\cite{Flouquet} In addition, this reduction depends on the magnitude of current, which indicates the onset of superconductivity. As shown in the inset of Fig.1, on annealing to improve the quality, the residual resistivity ratio (RRR) increases from RRR = 6 to 34 for the annealed sample. Nevertheless, the SC anomaly disappears. This is an underlying issue to be addressed on the basis of the microscopic measurements of the $^{73}$Ge-NQR spectrum and $1/T_1$ in order to clarify microscopically the non-Fermi-liquid-like behaviors  reported thus far and a possible signature of the onset of superconductivity.
 
 
\begin{figure}[h]
\centering
\includegraphics[width=5.2cm]{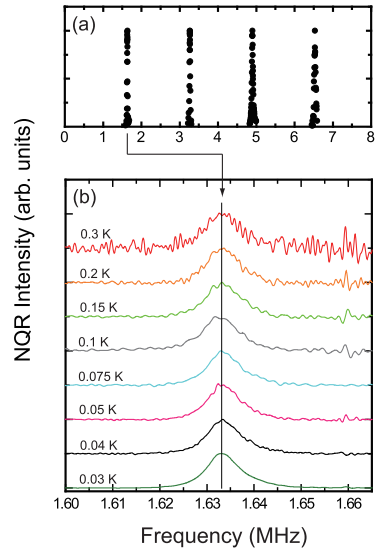}
\caption[]{\footnotesize (Color online) (a) $^{73}$Ge-NQR spectra at $T$ = 1.5 K (see text). (b) Temperature dependence of the $^{73}$Ge 1$\nu_{\rm Q}$-NQR spectrum in CeNi$_{2}$Ge$_2$. Solid line denotes the peak position.}
\end{figure}

Figure 2 (a) shows the $^{73}$Ge-NQR spectra at $T$ = 1.5 K, which consist of four equally spaced NQR spectra with  $\nu_{\rm Q}$ = 1.632 MHz and the asymmetry parameter $\eta$ = 0 due to the uniaxial symmetry.  Here, $\nu_{\rm Q}$ is defined as a parameter in the following Hamiltonian:
 
\begin{equation}
{\cal H}_{\rm Q}=\frac{h\nu{\rm _Q}}{6} (3I^2_{\rm z}-I^2 + \frac{1}{2}{\eta}(I^2_{+}+I^2_{-})),
\end{equation}
 
where
\begin{equation}
\nu{\rm _Q}=\frac{eQV_{zz}}{6I(2I+1)}
\end{equation}

and
\begin{equation}
{\eta}=\frac{V_{xx}-V_{yy}}{V_{zz}}.
\end{equation}
 
Note that the full width at half maximum (FWHM) for the 1$\nu_{\rm Q}$-NQR spectrum is quite sharp at FWHM = 8 kHz, confirming the high quality of the sample used in this study. The NQR spectra can indicate the emergence of a static magnetic order for CeNi$_2$Ge$_2$ from the splitting and/or the broadening of the 1$\nu_{\rm Q}$-NQR spectrum due to the appearance of an internal field, if any.  Since the FWHM of the 1$\nu_{\rm Q}$-NQR spectrum does not exhibit any change as shown in Fig. 2 (b), there is no evidence for a magnetic order and/or a structural change down to 0.03 K in the as-grown and annealed samples of CeNi$_2$Ge$_2$.

Next, we present the $T$ dependence of $1/T_1$ that reveals low-energy excitations relevant to non-Fermi-liquid-like behaviors and the possible onset of superconductivity in  CeNi$_2$Ge$_2$. Figures 3 (a) and 3 (b) show the typical data sets of the time dependence of nuclear magnetization of the 1$\nu_{\rm Q}$ transition at $T$ = 0.3 K and 0.04 K, respectively. They can be fitted by theoretical curves (solid lines) with a single $T_1$ component according to  Eq.(4) \cite{InT1} as indicated in Fig. 3. Note that the saturation pulse width is reduced below $T= 0.1$ K to avoid some heating effect caused by the application of a radio frequency field at low temperatures. 
 
\begin{eqnarray}
1-\frac{M(t)}{M_0}&=&\frac{1}{132}\mathrm{exp}\left(\frac{-3t}{T_1}\right)+\frac{45}{572}\mathrm{exp}\left(\frac{-10t}{T_1}\right)+\nonumber \\
& &\frac{49}{165}\mathrm{exp}\left(\frac{-21t}{T_1}\right)+\frac{441}{715}\mathrm{exp}\left(\frac{-36t}{T_1}\right).
\end{eqnarray}

\begin{figure}[h]
\centering
\includegraphics[width=5cm]{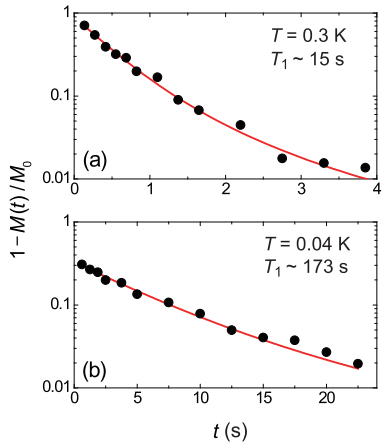}
\caption[]{\footnotesize (Color online) Recovery curves  of the nuclear magnetization of $^{73}$Ge  at  (a) $T$ = 0.3 K and (b) $T$ = 0.04 K in CeNi$_2$Ge$_2$. The red curves are fitted according to Eq.(4).}
\end{figure}
 
Figure 4 shows the $T$ dependence of $1/T_1T$ under zero field ($H=0$) (circle), $H$ = 0.03 T (triangle), and $H$ = 0.1 T (square) in the $T$ range of $0.04 - 150$ K. The $1/T_1T$ at $H=0$ increases as the temperature decreases to 0.4 K, thus revealing the growth of magnetic correlations. This result demonstrates that CeNi$_2$Ge$_2$ is very closely located to the AFM QCP. On the other hand, at temperatures lower than $T_{\rm FL}$ = 0.4 K, $1/T_1T$ is constant; this indicates the formation of the Fermi-liquid state. The development of magnetic correlations is corroborated by the recent inelastic neutron-diffraction measurements on a single crystal of CeNi$_2$Ge$_2$, these measurements have shown the presence of spin fluctuations with multiple $q$ structures.\cite{Kadowaki}  The result for the annealed sample in the inset  of Fig. 4 are almost the same as that of the as-grown sample above 0.1 K. 
 
\begin{figure}[h]
\centering
\includegraphics[width=7cm]{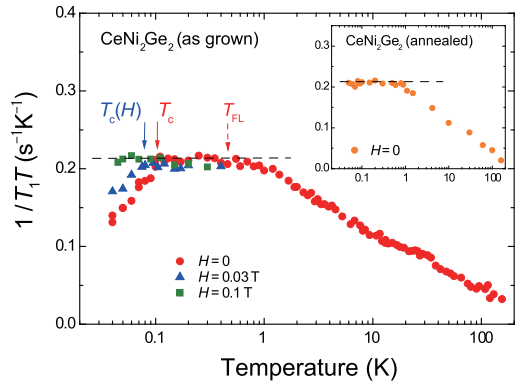}
\caption[]{\footnotesize (Color online) Temperature dependence of $1/T_1T$ for
CeNi$_{2}$Ge$_2$ (solid circle). Solid triangles and squares indicate the temperature dependence of $1/T_1T$ under a magnetic field of $H=0.03$ and 0.1 T, respectively. Dotted line indicates a $1/T_1T = const$ law. Solid and dashed arrows indicate $T_{\rm c}$ and $T_{\rm c}$($H$) and $T_{\rm FL}$ (see the text), respectively.}
\end{figure} 

For the as-grown sample, as shown in Fig. 4, a remarkable finding in the Fermi-liquid regime is that the $1/T_1T$ at $H=0$ slightly decreases below $T_{\rm c}^{\rm NQR} = 0.1$ K. Note that this temperature is lower than $T_{\rm c}^{\rm onset}$ = 0.2 K below which the resistance begins to decrease. In order to confirm whether this decrease in  $1/T_1T$ is associated with the onset of superconductivity, $1/T_1T$  was measured under external fields of $H$ = 0.03 and 0.1 T, as shown by the solid triangles and squares in the figure, respectively. The application of such tiny fields causes $1/T_1T$ to remain constant down to $T=0.07$ K and 0.04 K at $H=0.03$ T and 0.1 T, respectively. This result possibly provides evidence for the onset of superconductivity at $T_{\rm c}^{\rm NQR}=0.1$ K and 0.07 K at $H=0$ T and 0.03 T, respectively. 
Since $1/T_1T$ indicates a power-law like dependence without a coherence peak just below $T_{\rm c}$, the origin of superconductivity in CeNi$_2$Ge$_2$ seems to be unconventional. 

Notably, as shown in the inset of Fig. 4, $1/T_1T = const$ behavior is observed down to 0.04 K for the annealed sample.  Although the sample quality is improved by annealing, this SC anomaly in $1/T_1T$ as well as resistance disappears (see the inset of Fig. 1).
This is possibly because a slight unconventional superconductivity in CeNi$_2$Ge$_2$ is observed in the heavy Fermi-liquid state in the close vicinity of the AFM QCP.  As a result, the annealed sample seems to be apart from the SC QCP.   
\begin{figure}[h]
\centering
\includegraphics[width=5.7cm]{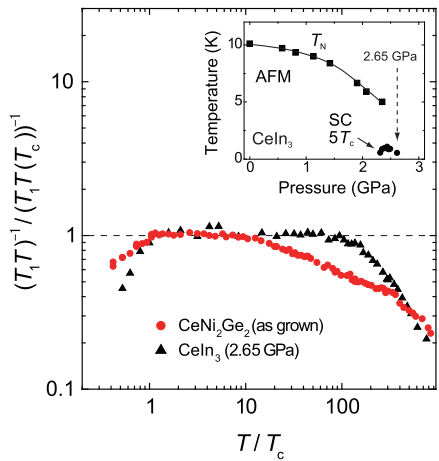}
\caption[]{\footnotesize (Color online) Temperature dependences of $1/T_1T$ of the as-grown sample of CeNi$_{2}$Ge$_2$ (solid circle) and CeIn$_3$ under a pressure of 2.65 GPa (solid triangle). The values of $1/T_1T$ and $T_{\rm c}$ are normalized  by  the values of $1/T_1T$ at $T_{\rm c}$ and $T_{\rm c}$, respectively. Dotted line indicates a $T_1T = const$ law. The inset shows a pressure-temperature phase diagram of antiferromagnetism and superconductivity in CeIn$_3$.\cite{Shinji,Shinji2} The dotted arrow corresponds to $P=2.65$ GPa.}
\end{figure}
 
In this context, it is likely that CeNi$_2$Ge$_2$ is closely located to the SC QCP. This is reinforced by the comparison of $1/T_1T$ for CeNi$_2$Ge$_2$ with that for pressure-induced unconventional superconductivity in CeIn$_3$ above $P_{\rm c}$.\cite{Shinji,Shinji2}  Figure 5 shows the $T$ dependences of $1/T_1T$ of the as-grown sample and CeIn$_3$ at $P$ = 2.65 GPa ($T_{\rm c}$ = 0.095 K).\cite{Shinji} The values of $1/T_1T$ and $T_{\rm c}$ are normalized by the values of $1/T_1T$ at $T_{\rm c}$ and $T_{\rm c}$, respectively. In CeIn$_3$, above $P_{\rm c}$ = 2.45 GPa, unconventional superconductivity is induced in the heavy Fermi-liquid state.\cite{Shinji2} 
Note that the superconductivity of CeIn$_3$ under $P=2.65$ GPa is revealed by the observations of zero resistance and SC diamagnetism.\cite{Shinji} A smaller reduction in  $1/T_1T$ for CeNi$_2$Ge$_2$ than for CeIn$_3$ below $T_{\rm c}$ could be attributed to the fact that the resistance of the as-grown sample does not become zero. These results are consistent with the fact that bulk superconductivity does not emerge but the SC coherence length remains finite over a short-range distance due to the closeness of the QCP for the SC order.  In another context, if the further increase in AFM fluctuations on cooling prevented the formation of the Fermi-liquid state in CeNi$_2$Ge$_2$, bulk superconductivity could be expected to occur due to the strong-coupling effect as demonstrated in the literatures.\cite{Kitaoka,Yashima} Thus, the onset of unconventional superconductivity in HF systems is closely related to the character of AFM spin fluctuations.

In conclusion, systematic studies of resistivity and $^{73}$Ge NQR in an as-grown sample of CeNi$_2$Ge$_2$ and its annealed one have revealed the development of magnetic correlations down to 0.4 K followed by the formation of a Fermi-liquid state below $T_{\rm FL}=0.4$ K. This very low Fermi energy is responsible for the non-Fermi-liquid-like behaviors discussed in the literatures.\cite{Steglich,Grosche,Aoki} A remarkable finding is that the significant reduction in $1/T_1T$ with a drop in resistance is associated with the onset of superconductivity below $T_{\rm c}=0.1$ K, this is because the application of a tiny magnetic field causes $1/T_1T$ to remain constant  without a reduction of $T_{\rm c}$ below 0.1 K. These results are due to the emergence of the unconventional SC fluctuations in association with the SC QCP. This is because zero resistance is not observed. As a result, these results are consistent with the fact that bulk superconductivity does not emerge but the SC coherence length remains finite over a short-range distance due to the closeness to the SC QCP.  We hope that the present works on CeNi$_2$Ge$_2$ will shed new light on the novel SC state in the heavy-Fermion compound CeNi$_2$Ge$_2$, which is in close proximity to the AFM QCP.

 This work was partially supported by a Grant-in-Aid for Creative Scientific Research (15GS0213) from the Ministry of Education, Culture, Sports, Science and Technology (MEXT) and the 21st Century COE Program (G18) by Japan Society of the Promotion of Science (JSPS).


\begin{thebibliography}{99} 
\bibitem{Kitaoka} Y. Kitaoka S. Kawasaki, T. Mito, Y. Kawasaki: J. Phys. Soc. Jpn. {\bf 74} (2005) 186 and references therein. 
 
\bibitem{Mathur} N. D. Mathur, F. M. Grosche, S. R. Julian, I. R. Walker, D. M. Freye, R. K. W. Haselwimmer, and G. G. Lonzarich: Nature {\bf 394} (1998) 39. 
\bibitem{Yuan} H. Q. Yuan, F. M. Grosche, M. Deppe, C. Geibel, G. Sparn, F. Steglich: Science {\bf 307} (2003) 2104.

\bibitem{Miyake} Y. Onishi and K. Miyake: J. Phys. Soc. Jpn. {\bf 69} (2000) 3955. 
 
 
\bibitem{Steglich} F. Steglich, B. Buschinger, P. Gegenwart, M. Lohmann, R. Helfrich, C. Langhammer, P. Hellmann, L. Donnevert, S. Thomas, A. Link, C. Geibel, M. Lang, G. Sparn, and W. Assmus: J. Phys.: Condens. Matter {\bf 8} (1996) 9909.
 
\bibitem{Grosche} F. M. Grosche, P. Agarwal, S. R. Julian, N. J. Wilson, R. K. W. Haselwimmer,
 S. J. S. Lister, N. D. Mathur, F. V. Carter, S. S. Saxena, and G. G. Lonzarich: cond-mat/9812133 (1998).

\bibitem{Aoki} Y. Aoki, J. Urakawa, H. Sugawara, H. Sato, T. Fukuhara, and K. Maezawa: J. Phys. Soc. Jpn. {\bf 66} (1997) 2993.
  

\bibitem{Sparn} G. Sparn, P.C. Canfield, P. Hellmann, M. Keller, A. Link, R.A. Fisher,
N.E. Phillips, J.D. Thompson, F. Steglich: Physica B {\bf 206 \& 207} (1995) 212.

\bibitem{Fukuhara} T. Fukuhara, S. Akamaru, T. Kuwai, J. Sakurai, K. Maezawa: J. Phys. Soc. Jpn. {\bf 67} (1998) 2084.


\bibitem{Knebel} G. Knebel, M. Brando, J. Hemberger, M. Nicklas, W. Trinkl, and A. Loidl: Phys. Rev. B {\bf 59} (1999) 12390.

  
\bibitem{Gegenwart} P. Gegenwart P. Hinze, C. Geibel, M. Lang, F. Steglich: Physica B {\bf 281 \& 282} (2000) 5.

\bibitem{Grosche2} F. M. Grosche, P. Agarwal, S. R. Julian, N. J. Wilson, R. K. W. Haselwimmer,
 S. J. S. Lister, N. D. Mathur, F. V. Carter, S. S. Saxena, and G. G. Lonzarich: J. Phys.: Condens. matter. {\bf 12} (2000) L533.
 
\bibitem{Flouquet} D. Braithwaite, T. Fukuhara, A. Demuer, I. Sheikin, S. Kambe,
 J.-P. Brison, K. Maezawa, T. Nakak, and J. Flouquet: J. Phys.: Condens. matter {\bf 12} (2000) 1339.

 
 
\bibitem{InT1} D.E.MacLaughlin, J. D. Williamson, and J. Butterworth: Phys. Rev. B {\bf 4} (1971) 60.


\bibitem{Kadowaki} H. Kadowaki, B. Fak, T. Fukuhara, K. Maezawa, K. Nakajima, M. A. Adams, S. Raymond, and J. Flouquet: Phys. Rev. B {\bf 68} (2003) 140402(R).

\bibitem{Shinji} S. Kawasaki, T. Mito, Y. Kawasaki, G.-q. Zheng, Y. Kitaoka, H. Shishido, S. Araki, R. Settai, and Y. \=Onuki: Phys. Rev. B {\bf 66} (2002) 054521.


\bibitem{Shinji2} S. Kawasaki, T. Mito, Y. Kawasaki, H. Kotegawa, G.-q. Zheng, Y. Kitaoka, H. Shishido, S. Araki, R. Settai, and Y. \=Onuki: J. Phys. Soc. Jpn. {\bf 73} (2004) 1647.

\bibitem{Yashima}
M. Yashima, S. Kawasaki, Y. Kawasaki, G.-q. Zheng, Y. Kitaoka, H. Shishido, R. Settai, Y. Haga, and Y. \=Onuki: J. Phys. Soc. Jpn. {\bf 73} (2004) 2073.

 
\end{thebibliography}
\end{document}